\documentclass{kapproc} 

\usepackage{procps} 
\usepackage[dvips]{graphicx}

\upperandlowercase
\kluwerbib 
\begin{document}

\articletitle[]{XMM-Newton discovery of O VII emission from warm gas in
clusters of galaxies}

\chaptitlerunninghead{XMM-Newton discovery of O VII}

\author{Jelle S. Kaastra\altaffilmark{1}, R. Lieu\altaffilmark{2}, 
T. Tamura\altaffilmark{1}, F.B.S. Paerels\altaffilmark{3} 
and J.W.A. den Herder\altaffilmark{1}}

\altaffiltext{1}{SRON National Institute for Space Research, Utrecht, The Netherlands}
\altaffiltext{2}{University of Alabama at Huntsville, USA}
\altaffiltext{3}{Columbia University, New York, USA}

\begin{abstract}
XMM-Newton recently discovered O~VII line emission from $\sim$ 2 million K gas
near the outer parts of several clusters of galaxies.  This emission is
attributed to the Warm-Hot Intergalactic Medium.  The original sample of
clusters studied for this purpose has been extended and two more clusters with a
soft X-ray excess have been found.  We discuss the physical properties of the
warm gas, in particular the density, spatial extent, abundances and temperature.
\end{abstract}


\section{Introduction}

Soft excess X-ray emission in clusters of galaxies was first discovered using
EUVE DS and Rosat PSPC data in the Coma and Virgo cluster (Lieu et al.
1996a,b).  It shows up at low energies as excess emission above what is expected
to be emitted by the hot intracluster gas, and it is often most prominent in the
outer parts of the cluster.  However, a serious drawback with the old data
(either EUVE or Rosat) concerns their spectral resolution, which does not exist
for the EUVE DS detector, and is very limited for the Rosat PSPC at low energies
where the width of the instrumental broadening causes a significant
contamination of the count rate by harder photons.

The forementioned reasons render it very difficult for firm conclusions about
the nature of the soft excess emission to be deduced from the original data
alone.  For example, already in the first papers it was suggested that the
emission may have a thermal origin, but is also consistent with it having a
power law spectrum caused by Inverse Compton scattering of the cosmic microwave
background on cosmic ray electrons (Sarazin \& Lieu 1998).

With the launch of XMM-Newton it is now possible to study the soft excess
emission with high sensitivity and with much better spectral resolution using
the EPIC camera's of this satellite.  The high resolution Reflection Grating
Spectrometer (RGS) of XMM-Newton has proven to be extremely useful in studies of
the central cooling flow region, but due to the very extended nature of the soft
excess emission, the RGS is not well suited to study this phenomemon.

\section{XMM-Newton observations}

XMM-Newton has by now observed a large number of clusters.  We investigated the
presence of soft excess emission in a sample of 14 clusters of galaxies.  This
work has been published by Kaastra et al.  (2003a).  In that paper the details
of the data analysis are given.  Briefly, much effort was devoted to subtracting
properly the time-variable soft proton background, as well as the diffuse X-ray
background.  We made a carefull assesment of the systematic uncertainties in the
remaining background, since a proper background subtraction has been one of the
contentious issues in the discussion around the discovery of the soft excess in
EUVE and Rosat data.  In a similar way, the systematic uncertainties in the
effective area and instrumental response of the EPIC camera's were carefully
assessed and quantified.  In the spectral fitting procedures, both the
systematic uncertainties in the backgrounds and the instrument calibration were
taken into account.  Spectra were accumulated in 9 concentric annuli between 0
and 15 arcmin from the center of the cluster.

The original sample used by Kaastra et al (2003a) has been extended to 21
clusters using archival data (see also Kaastra et al.  2003c).  These additional
clusters were analyzed in exactly the same way as the original 14 clusters.  The
spectra were initially analyzed using a two temperature model for the hot gas,
with the second temperature of the component fixed at half that of the first
component.  From experience with cooling flow analysis (Peterson et al.  2003;
Kaastra et al 2003b) we learnt that such a temperature parameterization is
sufficient to characterise fully the cooling gas in the cores of clusters, while
in the outer regions it is an effective method to take the effects of eventual
non-azimuthal variations in the annular spectra into account.  We note that the
temperature of the coolest "hot gas" component in all cases where we detect a
soft excess is much higher than the (effective) temperature of the soft excess.
This is due to the now well-known fact that the emission measure distribution of
the cooling flow drops off very rapidly.  In fact, our models for the cooling
flow predict no significant emission from O VII ions in the cooling plasma (at
least below the detection limit of XMM-Newton).

The presence of a soft excess in this sample of clusters was tested by formally
letting the Galactic absorption column density be a free parameter in the
spectral fitting.  Of the 21 clusters, 5 have apparent excess absorption.  All
these 5 clusters are located in regions where dust etc.  is important, or they
have a very compact core radius such that the temperature gradients in the core
are not fully resolved by XMM-Newton and therefore the spectra are highly
contaminated.

While the excess absorption in 5 clusters can be fully explained, the absorption
deficit in 7 of these clusters cannot be explained by uncertainties in the
calibration, background emission or foreground absorption, but only by the
presence of an additional emission component.  In fact, in several of the
clusters the best-fit column density is zero!

\begin{figure}
{\includegraphics[height=\hsize,width=0.30\vsize,angle=-90]{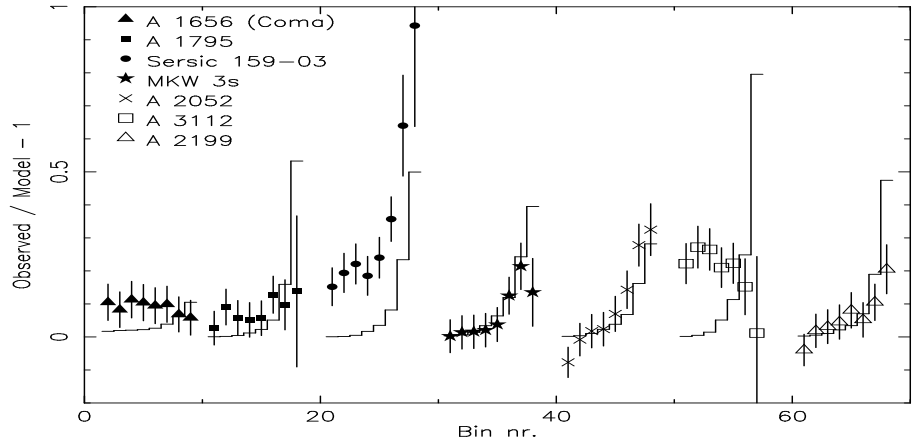}}
\caption{Soft excess in the 0.2--0.3~keV band as compared with a two temperature
model (data points with error bars).  The solid histogram is the predicted soft
excess based upon scaling the sky-averaged XMM-Newton background near the
cluster with the relative enhancement of the soft X-ray background derived from
the 1/4~keV PSPC images.  For the $k$th cluster the values for annulus $j$ are
plotted at bin number $10(k-1)+j$.}
\label{fig:soft}
\end{figure}

In Fig.~\ref{fig:soft} we show the soft excess in the 0.2--0.3~keV band for the
seven clusters with a significant soft component.  These clusters are Coma,
A~1795, S\'ersic~159$-$03, MKW~3s, A~2052 (see also Kaastra et al.  2003a);
A~3112 (see also Nevalainen et al.  2003 and Kaastra et al.  2003c) and A~2199
(paper in preparation).

The field of view of XMM-Newton is relatively small ($\sim$15 arcmin radius) and
therefore all these relatively nearby clusters fill the full field of view.  For
this reason, only a sky-averaged soft X-ray background (obtained from deep
fields) as well as the time variable soft proton background (which is relatively
small at low energies) were subtracted from the XMM-Newton data.  Using the
1/4~keV Rosat PSPC sky survey data (Snowden et al.  1995), maps at 40 arcmin
resolution were produced to estimate the average soft X-ray background in an
annulus between 1--2 degrees from the cluster.  Most of these seven clusters
show an enhanced 1/4~keV count rate in this annulus (as compared to the typical
sky-averaged 1/4~keV count rate).  This, combined with the decreasing density of
the hot gas in the outer parts of the cluster causes the apparent soft excess
with increasing relative brightness at larger radii in Fig.~\ref{fig:soft}.  The
figure shows complete consistency of the XMM-Newton data with the PSPC 1/4~keV
data in this respect for A~1795, S\'ersic~159$-$03, MKW~3s, A~2052 and A~2199.
We show below that in these clusters this large-scale soft X-ray emission is due
to thermal emission from the (super)cluster environment.

However, there is an additional soft component in Coma, A~3112, S\'ersic
159$-$03 and perhaps A~1795.  The fact that this component is above the
prediction from the large scale PSPC structures implies that its spatial extent
is at most 10--60 arcmin.  We shall return to this component in Sect.~5.

\section{Emission from the Warm-Hot Intergalactic Medium}

\begin{figure}
\resizebox{\hsize}{!}{\includegraphics[angle=0]{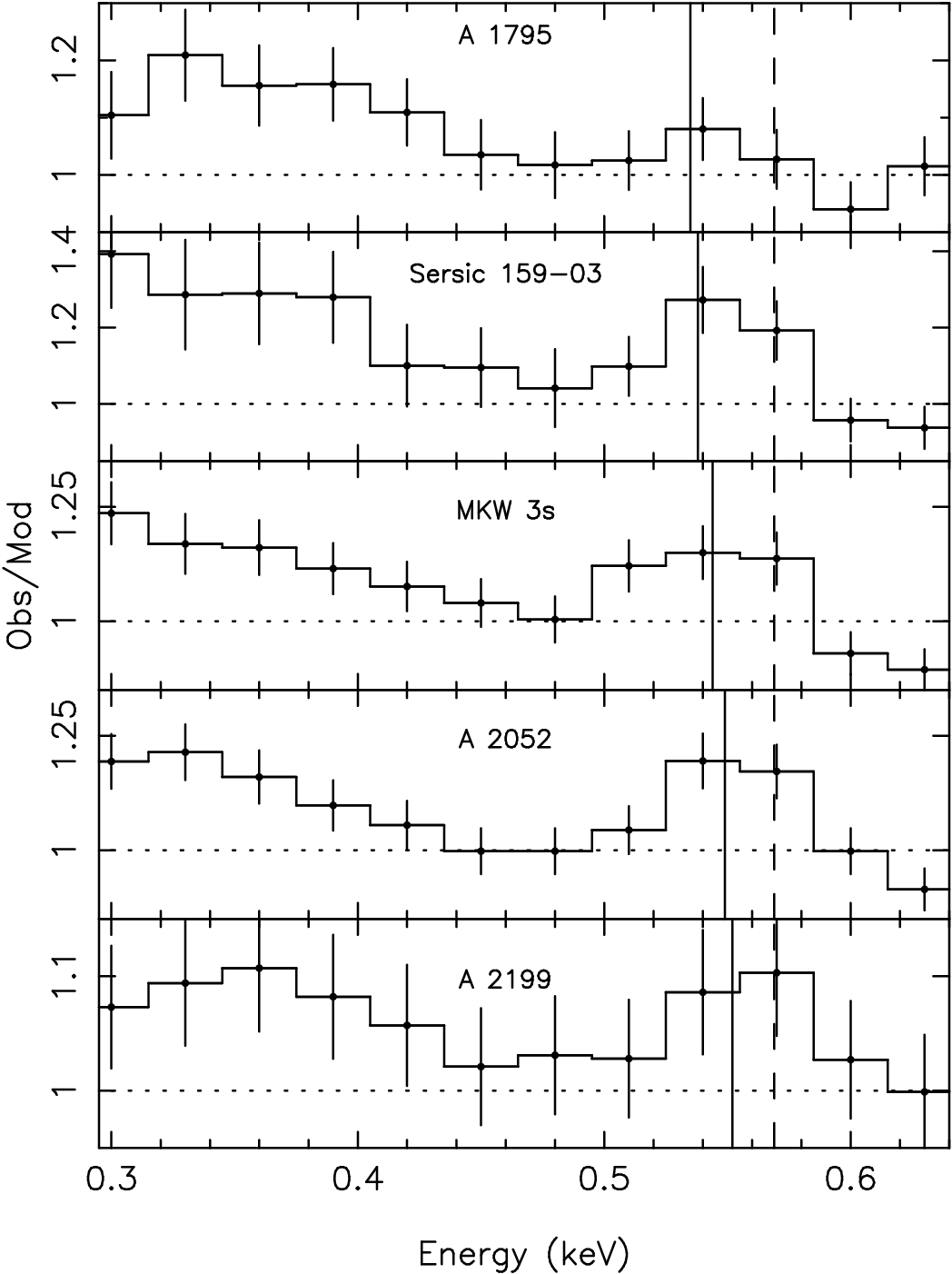}}
\caption{Fit residuals with respect to a two temperature model for the outer
4--12 arcmin part of six clusters.  The position of the O~VII triplet in the
cluster restframe is indicated by a solid line and in our Galaxy's rest frame by
a dashed line at 0.569~keV (21.80~\AA).  The fit residuals for all instruments
(MOS, pn) are combined.  The instrumental resolution at 0.5~keV is $\sim$60~eV
(FWHM).}
\label{fig:o7}
\end{figure}

In the previous section we found that five clusters show a soft excess in the
0.2--0.3~keV band at a spatial scale of at least 1--2 degrees, combining our
XMM-Newton spectra with PSPC 1/4~keV imaging.  It is not obvious a priori
whether this excess emission is due to emission from the cluster region or
whether it has a different origin, for example galactic foreground emission.
Here the spectral resolution of the EPIC camera's is crucial in deciding which
scenario is favoured.  In Fig.~\ref{fig:o7} we show the fit residuals of the fit
with two hot components and Galactic absorption only, in the outer 4--12 arcmin
region combining all three EPIC camera's.  The fit residuals show two distict
features:  a soft excess below 0.4--0.5~keV, and an emission line at
$\sim$0.56~keV.  This emission line is identified as the O~VII triplet, and
detailed spectral fitting shows that both phenomena (soft excess and O~VII line)
can be explained completely by emission from a warm plasma with a temperature of
0.2~keV (see Kaastra et al.  2003a for more details).  Thus, the soft excess has
a thermal origin.  Moreover, the centroid of the O~VII triplet (which is
unresolved) agrees better with an origin at the redshift of the cluster than
with redshift zero.  This clearly shows that the thermal emission has an origin
in or near the cluster, although a partial contribution from Galactic foreground
emission cannot be excluded in all cases.  We also note that in A~1795 and
A~2199 the O~VII line is relatively weak and needs more confirmation.

Taking this additional soft thermal component into account in the spectral
fitting yields fully acceptable fits.  In fact, in the energy band below 1~keV,
the soft component contributes 20--40~\% of the X-ray flux of the outer (4--12
arcmin) part of the cluster!

We identify this component as emission from the Warm-Hot Intergalactic Medium
(WHIM).  Numerical models (for example Cen \& Ostriker 1999, Fang et al.  2002)
show that bright clusters of galaxies are connected by filaments that contain a
significant fraction of all baryonic matter.  Gas falls in towards the clusters
along these filaments, and is shocked and heated during its accretion onto the
cluster.  Near the outer parts of the clusters the gas reaches its highest
temperature and density, and it is here that we expect to see most of the X-ray
emission of the warm gas.

\section{Properties of the warm gas}

The temperature of the warm gas that we find for all our clusters is 0.2~keV.
The surface brightness of the warm gas within the field of view of the
XMM-Newton telescopes is approximately constant, with only a slight enhancement
towards the center for some clusters (a stronger increase towards the center for
S\'ersic 159$-03$ is discussed in the next section).  In Table~1 we list the
central surface brightness $S_0$ as estimated from our XMM-Newton data,
expressed as the emission measure per solid angle.  We use
$H_0=70$~km\,s$^{-1}$\,Mpc$^{-1}$ throughout this paper.  Using the known
angular distance to the cluster we estimate the same quantity in units of
m$^{-5}$ (see also Table~1).  We then use a simplified model for the geometry of
the emitting warm gas, namely a homogeneous sphere with uniform density.  The
radius $R$ of this sphere is estimated from the radial profile of the Rosat PSPC
1/4~keV profile around the cluster, and is also listed in Table~1.  We find
typical radii of 2--6~Mpc, i.e.  the emission occurs on the spatial scale of a
supercluster.  From this radius and the emission measure, the central hydrogen
density $n_{\mathrm H}(0)$ is estimated.  We find typical densities of the order
of 50-150~m$^{-3}$.  Assuming a different density profile (for example
$n(r)=n(0)[1+(r/a)^2]^{-1}$ for $r<R$ and $n=0$ for $r>R$) yields central
densities that are only 20--50~\% larger.  These densities are 200--600 times
the average baryon density of the universe.  We also estimate the total hydrogen
column density, which is typically $1.6-2.8\times 10^{25}$~m$^{-2}$.  Using then
the measured oxygen abundances from our XMM-Newton data (essentially determined
by the ratio of the O~VII triplet to the soft X-ray excess), which are typically
0.1 times solar, we then derive total O~VII column densities of the order of
$0.4-0.9\times 10^{21}$~m$^{-2}$.  These column densities and the typical sizes
of the emitting regions are similar to those as calculated for the brightest
regions in the simulations of Fang et al.  (2002).  We have taken here a solar
oxygen abundance of $8.5\times 10^{-4}$ and an O~VII fraction of 32~\%,
corresponding to a plasma with a temperature of 0.2~keV.

\begin{table}
\caption{Properties of the warm gas}
\begin{tabular*}{\textwidth}{@{\extracolsep{\fill}}lccccc}
\hline
Parameter & A~1795 & S\'ersic 159$-03$ & MKW~3s & A~2052 & A~2199 \\
\hline
Redshift & 0.064 & 0.057 & 0.046 & 0.036 & 0.030 \\
Scale (kpc/arcmin) & 71 & 64 & 52 & 41 & 35 \\
$S_0$\,$^a$ & 49$\pm$37  & 71$\pm$41 & 84$\pm$18 & 73$\pm$15 & 28$\pm$13 \\
$S_0$\,$^b$ & 10   & 18  & 33  & 46  & 24 \\
$R$ (arcmin) & 80  & 60  & 60 & 80 & 60 \\
$R$ (Mpc)    & 5.7 & 3.8 & 3.1 & 3.3 & 2.1 \\
$n_{\mathrm H}(0)$ (m$^{-3}$) & 45 & 80 & 120 & 140 & 120 \\
$N_{\mathrm H}(0)$ ($10^{24}$~m$^{-2}$) & 16 & 19 & 23 & 28 & 16 \\
Abundance O & 0.08$\pm$0.05 & 0.08$\pm$0.03 & 0.09$\pm$0.03 & 0.12$\pm$0.03 & 
         0.16$\pm$0.05 \\
$N_{\mathrm O VII}(0)$ ($10^{20}$~m$^{-2}$) & 4 & 4 & 6 & 9 & 7 \\
$M_w$ ($10^{15}$~M$_{\odot}$) & 1.2 & 0.6 & 0.5 & 0.7 & 0.2 \\
$M_{\mathrm A}$ ($10^{15}$~M$_{\odot}$) & 1.1 & 0.5 & 0.5 & 0.4 & 0.6 \\
\hline
\end{tabular*}
\begin{tablenotes}
$^a$Surface brightness, expressed as emission measure per
solid angle ($10^{68}$~m$^{-3}$arcmin$^{-2}$).

$^b$Surface brightness in $10^{26}$~m$^{-5}$.
\end{tablenotes}
\end{table}

Finally, we determined the total mass of the warm gas ($M_w$, Table~1).  This
mass is for most clusters comparable to the total cluster mass $M_A$ within the
Abell radius (2.1~Mpc for our choice of $H_0$) as derived by Reiprich \&
B\"ohringer (2002) for the same clusters.

We make here a remark on MKW~3s and A~2052.  These clusters are separated by
only 1.4 degree and both belong to the southernmost extension of the Hercules
supercluster (Einasto et al.  2001).  A~2199, at 35 degrees to the North, is at
the northernmost end of the same supercluster.  The redshift distribution of the
individual galaxies in the region surrounding MKW~3s ($z=0.046$) and A~2052
($z=0.036$) shows two broad peaks centered around the redshifts of these
clusters, but galaxies with both redshifts are found near both clusters.
Therefore this region has a significant depth (43~Mpc) as compared to the
projected angular separation (4.3~Mpc).  The relative brightness of the warm gas
near these clusters (as seen for example from the value of $S_0$) is then
explained naturally if there is a filament connecting both clusters.  In that
case we would see the filament almost along its major axis.

\section{Non-thermal emission?}

Apart from the large scale, extended emission from the warm gas some clusters
also exhibit a centrally condensed soft excess component (Fig.~3).  In MKW~3s,
A~2052 and A~2199 this central enhancement is relatively weak.  It could be a
natural effect of the enhanced filament density close to the cluster centers.
In A~1795 and S\'ersic 159$-03$ the enhancement is much larger.  It is unlikely
that this emission component for the latter two clusters also originates from
projected filaments in the line of sight - the high surface brightness would
necessitate filaments of length far larger than a cluster's dimension.  Another
possibility is warm gas within the cluster itself.  In order to avoid the rapid
cooling which results from this gas assuming a density sufficient to secure
pressure equilibrium with the hot virialized intracluster medium, it should be
be magnetically isolated from the hot gas.  Yet another viable model for the
central soft component is non-thermal emission.  We note that the soft excess in
the center of Coma and A~3112 also possibly has a non-thermal origin, as there
is no clear evidence for oxygen line emission in their spectra.  Clearly, deeper
spectra and in particular a higher spectral resolution is needed to discriminate
models.  At this conference, several new mission concepts have been presented
that may resolve these issues in the near future.

\begin{figure}
{\includegraphics[height=\hsize,width=0.55\vsize,angle=-90]{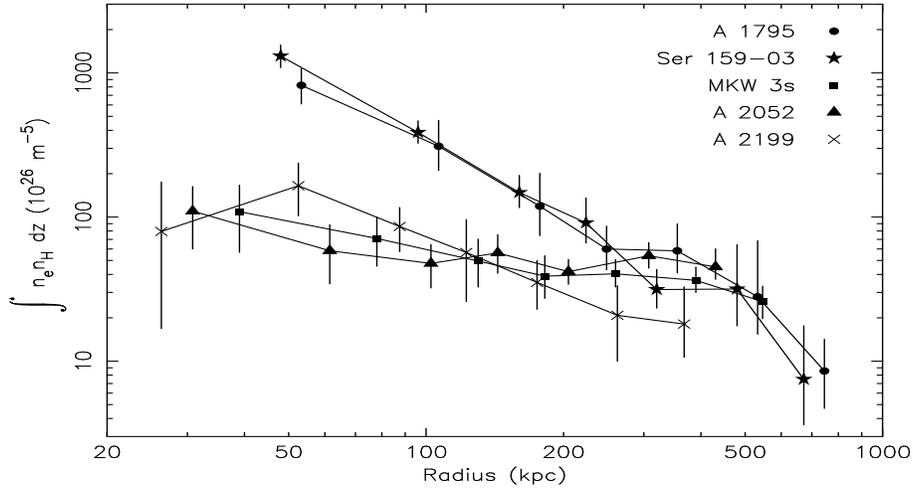}}
\caption{Emission measure integrated along the line of sight for five
clusters of galaxies.}
\label{fig:sb}
\end{figure}

\begin{acknowledgments}
This work is based on observations obtained with XMM-Newton, an ESA science
mission with instruments and contributions directly funded by ESA Member States
and the USA (NASA).  SRON is supported financially by NWO, the Netherlands
Organization for Scientific Research.
\end{acknowledgments}

\end{document}